\documentclass[aps,pre,showpacs,superscriptaddress,twocolumn,amsmath,amssymb]{revtex4}

\usepackage{graphicx}
\usepackage{epsfig}
\usepackage{bm}
\usepackage{bbm}

\usepackage{epic}
\usepackage{eepic}
\usepackage{pifont}
\usepackage[usenames]{color}

\usepackage{nicefrac}
\def\vec#1{\mathbf{#1}}

\input epsf
\input rotate
\usepackage{graphicx}

\newcommand{\be}{\begin{equation}}
\newcommand{\ee}{\end{equation}}
\newcommand{\bea}{\begin{eqnarray}}
\newcommand{\eea}{\end{eqnarray}}

\newcommand{\fra}[2]{\hbox{${#1\over #2}$}}

\newcommand{\comment}[1]{}
\newcommand{\sq}{{\hbox{\tiny sq}}}
\newcommand{\hex}{{\hbox{\tiny hex}}}
\newcommand{\con}{{\hbox{\tiny C}}}
\newcommand{\non}{{\hbox{\tiny NC}}}

\begin{document}

\title{Phase instability and coarsening in two dimensions}

\author{Chaouqi Misbah}
\email{chaouqi.misbah@ujf-grenoble.fr}
\affiliation{Laboratoire de Spectrom\'etrie Physique, UMR, 140 avenue de la
physique, Universit\'e Joseph Fourier Grenoble, and CNRS, 38402 Saint Martin
d'Heres, France}

\author{Paolo Politi}
\email{paolo.politi@isc.cnr.it}
\affiliation{Istituto dei Sistemi Complessi, Consiglio
Nazionale Delle Richerche, Via Madonna Del Piano 10, 50019, Sesto
Fiorentino, Italy}

\date{\today}
\begin{abstract}
Instabilities and pattern formation is the rule in nonequilibrium
systems. Selection of a  persistent lengthscale, or coarsening
(increase of the lengthscale with time) are the two major
alternatives.
When and under which conditions one dynamics prevails over the other
 is a longstanding problem, particularly beyond one dimension.
It is shown that the challenge can be  defied in two dimensions,
using the concept of phase diffusion equation.
We find that coarsening is related to the $\lambda$---dependence
of a suitable phase diffusion coefficient, $D_{11}(\lambda)$, depending
on lattice symmetry and conservation laws. These results
are exemplified analytically on prototypical nonlinear
equations.
\end{abstract}

\pacs{05.70.Ln, 05.45.-a, 64.60.Ht, 68.43.Jk}



\maketitle

{\it Introduction}.---Pattern formation (or morphogenesis) is
abundant in nature, both in inanimate and living systems.
Examples~\cite{book_Ball} are encountered in many branches of
science: physics (e.g. sand ripples), chemistry (chemical spots,
reminiscent of those on animal skins, e.g.  jaguar), biology (asters
of macromolecules during cell division), and so on. Patterns arise
often due to the loss of stability of an initially structureless
state. The diversity and richness of morphogenesis  is concomitant
to the nonlinear and nonequilibrium nature of these systems.

A  classification of dynamics and patterns that prevail for a given
nonlinear system  is a  challenging task.
For example, given a   system described by non linear equations, it
is not obvious to state a priori whether one would assist to the
selection of a pattern with a specific lengthscale during time, or
rather coarsening (increase of lengthscale with time) would prevail.

Coarsening is also the primary scenario in phase separation
processes~\cite{review_Bray} and it attracts a continuous
interest~\cite{coarsening_with_gravity,Bray_Cugliandolo_PRL,%
Bray_Cugliandolo_PRE,Lifshitz-Safran,Watson,Politi_MisbahI,%
Kohn_Yan,driven_granular_media} motivated by the demand for a deeper
understanding. This goal may be attained following two main
directions: either investigating rigorous solutions to specific
models~\cite{Watson,Bray_Cugliandolo_PRL}, or developing   general
approaches which are valid for wide classes of models/equations.
Regarding the second strategy, the only example we are aware of is
due to A.~Bray and collaborators~\cite{review_Bray,Bray_Rutenberg},
who analyzed some models which are derivable from a potential: comparing
the global rate of energy change with  the energy dissipation, the
temporal growth law for the typical lengthscale, $\lambda(t)$, could
be derived. 
The present  study can be seen as complementary, in two respects.
First, we do not impose to  the equations to be derivable from a
potential. Second, we do not assume that coarsening does occur, we
are able to determine if and when it occurs, instead. Note that our study does not address the other  kind of coarsening  which consists in an increase of a domain with a certain orientation (or pattern),  and/or lengthscale, at the expense of another domain \cite{Cross1995}.

A few years ago we established a coarsening criterion in one
dimension (1D) \cite{Politi_MisbahI}, for certain classes of nonlinear
equations, generalizations of the celebrated models $A$ and $B$ of
the dynamics~\cite{Hohenberg_Halperin}. The criterion demands to
analyse the stationary periodic states, which are found to solve
Newton's equations where the spatial variable plays the role of time
and the fictitious particle oscillates in a potential well of
arbitrary shape (different equations give rise to different
potentials). The oscillation period of the particle corresponds, in
the mechanical analogy, to the wavelength $\lambda$ of the periodic
steady state and it is a function of the amplitude $A$,
$\lambda=\lambda(A)$. The criterion simply states that coarsening
occurs if and only if $d\lambda/dA > 0$.

Our analysis anticipated the existence of an ``interrupted coarsening"
scenario, which has recently been invoked~\cite{Andreotti}
in the context of wind driven sand dunes. It is clear that these
physical phenomena (as well as, e.g., mound coarsening
at the nanoscale in crystal growth~\cite{mounds}) demand for the
construction of a full two-dimensional approach,
which is the basic goal of the present paper.

However, in our former 1D criterium,
the spatial variable plays the role of time (see above).
Therefore, it is obvious that  this
concept is limited to one spatial dimensional systems and
extensions of the criterion to higher dimensions seemed to present
a significant challenge.
It is shown here that the challenge to find general criteria for
understanding if a two dimensional (2D) model shows coarsening or
not can be defied, following the concept of phase diffusion
equation~\cite{book_Hoyle}. Before discussing the details, we give a
simple overview of the method and results.

{\it The method}.---A perfectly periodic steady state is defined
through a pair of wave wectors $\vec q_1, \vec q_2$ and it is a
function of the phases $\phi_{1,2}=\vec q_{1,2}\cdot \vec x$.
Perturbations make $\vec q_{1,2}$ acquire slow dependencies on time
and space, $\vec q_i=\vec q_i(\vec X,T)$. The dynamical response to
perturbations is described by phase equations, which have, at the
lowest order, the form of linear diffusion equations: \be
\begin{split}
\partial_T \phi_1 (\vec X,T) &= D^{11}_{11} {\partial^2 \phi_1\over\partial X_1^2}
+ D^{12}_{11} {\partial^2 \phi_2\over\partial X_1^2} +
+ D^{11}_{22} {\partial^2 \phi_1\over\partial X_2^2} +
\cdots \\
\partial_T \phi_2 (\vec X,T) &= D^{21}_{11} {\partial^2 \phi_1\over\partial X_1^2}
+ D^{22}_{11} {\partial^2 \phi_2\over\partial X_1^2} +
+ D^{21}_{22} {\partial^2 \phi_1\over\partial X_2^2} +
\cdots
\end{split}
\ee

Setting $\phi_{1,2}=\phi^0_{1,2}\exp(\omega T + i\vec K\cdot\vec
X)$, standard Fourier analysis allows to determine the
stable/unstable character of phase dynamics. Coarsening is related
to instability, which is signaled by a positive $\omega$. The
requirement $\omega>0$ for some $\vec K$ implies a condition on the
diffusion coefficients $D^{ij}_{kl}$, which are functions of steady
state properties only. The central result is that the condition
$\omega>0$ finally writes as $d{\cal A}/d\lambda > 0$, where ${\cal
A}$ is some function of the stationary periodic solution having
wavelength $\lambda$.

In one dimension, ${\cal A}$ is shown to correspond to the amplitude
of the stationary solution  \cite{Politi_MisbahI}. In two
dimensions, ${\cal A}$ takes different expressions, depending on the
underlying symmetry of the periodic stationary solutions, and it
dees do not seem to have a simple physical interpretation.
Nevertheless, the condition $d{\cal A}/d\lambda > 0$ does have a
simple reading: coarsening occurs if and only if ${\cal A}$ is an
increasing function of $\lambda$. That is to say the coarsening
criterion can still be established only upon inspecting {\it
steady-state solutions}. Even more importantly, the coarsening law,
$\lambda(t)$, which describes the time dependence of the typical
size of the pattern, is obtained through a  relation of the form $D
(\lambda) t \sim \lambda^2$ (or equivalently $\omega\sim
1/t$), where $D$ is a typical diffusion constant. In the limit of
large time, a power law behavior, $\lambda \simeq t^n$, is expected.

We shall exemplify the method on classical model equations. For
instance, we show that  $n={1\over 2}$ for the non-conserved real
Ginzburg-Landau equation (model A), and $n={1\over 3}$ for the
conserved Cahn-Hilliard equation (model B). As will be recognized,
the interesting message is that the methodology does not evoke
whether the model equation are variational or not. Thus the study
should   work with any other equation, be it of potential (i.e.
variational) nature or not.

{\it Generalized Ginzburg-Landau type models}.---We now apply and
discuss our method for a nonconserved class of equations, \be
\partial_t u = A(u) + \nabla^2 u ,
\label{gGL} \ee which reduces, for $A(u)=u-u^3$, to the famous real
Ginzburg-Landau (or Allen-Cahn) equation. This equation is known to
exhibit perpetual coarsening, with $n={1\over
2}$~\cite{review_Bray}. Here we only require $A(u) \approx u$ for
small $u$, otherwise   $A(u)$ can be
any function of $u$.

A steady periodic state $u_0(\vec x)$ depends on the fast spatial
variables $\vec x = (x_1,x_2)$, or equivalently  $\phi_{1,2} = \vec
q_{1,2}\cdot\vec x$ which are the fast (constant) phases. Phase
modes are studied by perturbing the perfectly periodic  steady
state. The perturbations of interest  depend on  slow spatial and
temporal variables. The slow character of relevant perturbations
reflects the property of the  Goldstone mode (if $u_0({\bf x})$ is a
periodic solution, so is $u_0({\bf x+R_0})$),
which has
infinite relaxation time and is a quasi-dangerous mode, making
slowly varying  perturbations to be most persistent. Let us encode
the slow modulation by a dimensionless parameter $\epsilon \ll 1$.
We define slow spatial and time scales as \be X_1 = \epsilon x_1,~~~
X_2 = \epsilon x_2,~~~ T = \epsilon^2 t, \ee where the factor
$\epsilon^2$ is an ansatz dictated by the fact that phase
rearrangement occurs via diffusion. We  introduce the slow phases
$\psi_{1,2}=\epsilon\phi_{1,2}$, so that $\vec q_i = \nabla_{\vec
x}\phi_i = \nabla_{\vec X}\psi_i$, where $\vec X = (X_1,X_2) =
\epsilon \vec x$.

In general terms, at order $\epsilon$ we can write~\cite{Politi_MisbahI}
\bea
\partial_t &=& \epsilon \left[
{\partial\psi_1\over\partial T} \partial_{\phi_1} +
{\partial\psi_2\over\partial T} \partial_{\phi_2} \right] \label{eq_T}\\
\nabla_{\vec x} &=& \vec q_1 \partial_{\phi_1} +\vec q_2 \partial_{\phi_2}
+\epsilon \nabla_{\vec X} . \label{eq_X}
\eea

The above expressions for the derivative, along with the standard
expansion $ u = u_0 + \epsilon u_1$, are  substituted into
Eq.~(\ref{gGL}). The zeroth order problem yields the differential
equation obeyed by the two-dimensional profile of steady states, \be
{\cal N}[u_0] \equiv A(u_0) + \nabla_0^2 u_0 = 0 \ee where the
subscript in the $\nabla^2$ operator is taken to mean that it is
evaluated for $\epsilon=0$ in Eq.~(\ref{eq_X}).

The next order in $\epsilon$ has the form~\cite{Politi_MisbahII,book_Hoyle}
\be
{\cal L}[u_1] = g(u_0,\partial_T\psi_i,\partial_{X_i X_j}\psi_k) ,
\label{1st_order}
\ee
where ${\cal L}={\cal L}_0=(A'(u_0) + \nabla_0^2)$
is the linear operator obtained as Frech\'et derivative
of ${\cal N}$, and
\begin{widetext}
\be
g = {\partial \psi_1\over\partial T}
{\partial u_0 \over\partial \phi_1} +
{\partial \psi_2\over\partial T}
{\partial u_0 \over\partial \phi_2}
- \left[
2 ( \vec q_1 \partial_{\phi_1} + \vec q_2 \partial_{\phi_2} )
\cdot \nabla_{\vec X} u_0
+ (\nabla_{\vec X}\cdot \vec q_1) \partial_{\phi_1} u_0
+ (\nabla_{\vec X}\cdot \vec q_2) \partial_{\phi_2} u_0
\right] .
\label{eq_g}
\ee
\end{widetext}
The expression for $g$ can be rewritten as \be g= (\partial_T
\psi_1) v_1 + (\partial_T \psi_2) v_2 -2q_{ki}(\psi_l)_{ij}
{\partial v_k\over
\partial q_{lj}} - (\psi_l)_{ii} v_l \ee where
$v_i=\partial_{\phi_i}u_0$, $q_{ki}$ is the $i-$th component of
$\vec q_k$ and $(\psi_l)_{ij}=\partial^2 \psi_l/\partial X_i\partial
X_j$.

In order to avoid secular terms in Eq.~(\ref{1st_order}),
if the homogeneous equation
${\cal L}_0^\dagger [w]=0$ has a nonvanishing solution, $w$ must be
orthogonal to $g$~\cite{book_Nayfeh}, $\langle w,g\rangle=0$
(the precise definition of the inner product is given below).

There exit two nonvanishing solutions $w_1,w_2$ in two dimensions,
resulting thus in two solvability conditions. Since ${\cal
L}_0^\dagger={\cal L}_0$ for Eq.~(\ref{gGL}), we easily get
$w_i=v_i$. This  leads to the sought-after phase diffusion
equations, \bea \langle v^2\rangle
\partial_T \psi_1 + \langle v_1v_2\rangle\partial_T\psi_2
&=& \sum_{ijl} (\psi_l)_{ij} A^l_{ij} \\
\langle v_1v_2\rangle \partial_T \psi_1 + \langle v^2\rangle\partial_T\psi_2
&=& \sum_{ijl} (\psi_l)_{ij} B^l_{ij}
\eea
where $\langle v^2\rangle= \langle v_1^2\rangle =\langle v_2^2\rangle$ and
\bea
A^l_{ij} &=& 2\sum_k q_{ki}\langle v_1{\partial v_k\over\partial q_{lj}}\rangle
+\delta_{ij}\langle v_1v_l\rangle \\
B^l_{ij} &=& 2\sum_k q_{ki}\langle v_2{\partial v_k\over\partial q_{lj}}\rangle
+\delta_{ij}\langle v_2v_l\rangle .
\eea

There are  five Bravais lattices in 2D (oblique, rectangular,
centered rectangular, hexagonal, and square). We have performed
explicitly the calculation for squares, hexagons and rectangles. Here, we
shall focus on the first two symmetries.
For square symmetry, $\vec q_1=q(1,0)$, $\vec
q_2=q(0,1)$ and $\langle v_1v_2\rangle=0$. Finally, we get \be
\begin{split}
{\partial
\psi_1\over\partial T} &=
 D_{11} {\partial^2\psi_1\over\partial X_1^2} +
 {\partial^2\psi_1\over\partial X_2^2} +
 D_{12} {\partial^2\psi_2\over\partial X_1X_2} \\
{\partial \psi_2\over\partial T} &=
 {\partial^2\psi_2\over\partial X_1^2} +
D_{11} {\partial^2\psi_2\over\partial X_2^2} + D_{12}
{\partial^2\psi_1\over\partial X_1X_2} \label{pd}
\end{split}
\ee with $D_{11}=D_{11}^\sq =\partial_q(q\langle v^2\rangle)/\langle
v^2\rangle$ and $D_{12}=D_{12}^\sq =D_{11}^\sq-1$. A linear
stability analysis is performed by setting $\psi_{1,2} =
\psi_{1,2}^0\exp(\omega T +i\vec K \cdot\vec X)$ in Eqs.~(\ref{pd}),
which yields \be \omega^2 + \omega (1+D_{11})K^2 + D_{11} K^4 =0 \ee
whose roots are $\omega_1=-K^2$ and $\omega_2=-D_{11}K^2$.

While $\omega_1<0$, the sign of $\omega_\non=\omega_2$ depends on the sign of
$\partial_q {\cal A} =\partial_q (q\langle v^2\rangle)$. Therefore,
$\omega_2$ is positive, implying phase instability, if and only if
${\cal A}= q\langle v^2\rangle$ is an increasing function of the
wavelength $\lambda$.

For hexagonal symmetry, $\vec q_{1,2}=
q(\fra{1}{2},\pm\fra{\sqrt{3}}{2})$ and $\psi_{1,2}$ must be
linearly combined in order to get Eqs.~(\ref{pd}), with a different
\be D_{11}=D_{11}^\hex = {\partial_q (q\langle v^2\rangle^2)\over
\langle v^2\rangle^2} \ee albeit the relation $D_{12}^\hex =
D_{11}^\hex -1$ still holds.

{\it Cahn-Hilliard type models}.---Let us now consider a conserved
equation, a generalized form of the well-known Cahn-Hilliard equation,
\be
\partial_t u = - \nabla^2 \left( A(u) + \nabla^2 u \right) .
\label{gCH}
\ee

We only provide the result. We formally obtain
Eq.~(\ref{1st_order}), but now
${\cal L} =
-\nabla_0^2 {\cal L}_0$
and
$ g = (\partial_T \psi_1) v_1 + (\partial_T \psi_2) v_2
+\nabla_0^2\nabla_1^2 u_0$
where $\nabla_1^2 u_0$ is a shorthand
notation for the expression in square parentheses in
Eq.~(\ref{eq_g}). ${\cal L}$ is not self-adjoint, because ${\cal
L}^\dagger = -{\cal L}_0\nabla_0^2$ and its kernel $w_i$ is such
that ${\cal L}^\dagger w_i =0=-{\cal L}_0 \nabla_0^2 w_i$, so that
$\nabla_0^2 w_i = v_i = {\partial
u_0\over\partial\phi_i}$.

The orthogonality conditions $\langle w_i g\rangle=0$ are a bit more
challenging to analyse, and will be discussed elsewhere. Suffice it
here to say that the relevant eigenvalue reads \be \omega_\con =
I_\con \omega_\non\,,\ee where $\omega_\con$ and $\omega_\non$ are the
eigenvalues associated with the unstable mode for the Conserved and
Non Conserved versions of a given model, and $I_\con = \langle v^2
\rangle / \langle wv \rangle$ is the factor taking into account the
conservation law.

{\it The coarsening exponent}.---The coarsening law is determined on
the basis of the relation $|\omega (q)| \sim 1/t$, where
$\omega=\omega_\non = -D_{11}q^2$ or $\omega=\omega_\con = I_\con
\omega_\non$. A major role is played by the quantity \bea \langle
v^2 \rangle &=& {1\over (2\pi)^2} \int_0^{2\pi} d\phi_1
\int_0^{2\pi} d\phi_2 \left( {\partial
u_0\over\partial\phi_1}\right)^2
\nonumber\\
&=& {1\over (2\pi)^2} \int_0^\lambda dx_1 \int_0^\lambda dx_2 \left(
{\partial u_0\over\partial x_1}\right)^2 . \label{inner} \eea

Consider first the Ginzburg-Landau (nonconserved) equation. In this
case, $u_0$ is a constant except along domain walls,
so $\langle v^2 \rangle$ scales linearly with $\lambda$. More
precisely, $\langle v^2 \rangle = a/q +b$. Reporting this expression
into $D_{11}$ yields $n=\fra{1}{3}$ for square symmetry and
$n=\fra{1}{2}$ for hexagonal symmetry. Preliminary analysis for
rectangular symmetry leads to $n=\fra{1}{2}$ as well. In general had
$\langle v^2 \rangle$ scaled as $1/q^\alpha$ ($\alpha
> 1$) we would have obtained
 $n=\fra{1}{2}$ for all symmetries. Therefore, GL scaling
($\langle v^2 \rangle \sim 1/q$) obtained for the  square symmetry
seems to be singular: all other cases provide $n=\fra{1}{2}$.

The computation of the coarsening exponent for conserved models
requires  estimation of $I_\con = \langle v^2 \rangle / \langle
wv\rangle$. Since $\nabla^2 w = v$, we infer (after integration by
parts) $I_\con \sim \langle v^2 \rangle / \langle u_0^2 \rangle \sim
q$. Therefore, for the standard Cahn-Hilliard model, we get
$n=\fra{1}{4}$ for square symmetry and $n=\fra{1}{3}$ for other
symmetries. Table~\ref{tab_n} summarizes the results. Note that
standard GL and CH refer to $A(u)=u-u^3$, so that $\langle
v^2\rangle\sim q^{-1}$. In contrast, modified GL
equation means $\langle v^2\rangle\sim q^{-\alpha}$, with $\alpha
> 1$.

\begin{table}
\centering
\begin{tabular}{|r|c|c|}
\hline
 & ~~square~~ & ~~other symmetries~~ \\
\hline
standard GL & 1/3 & 1/2 \\
\hline
modified GL & 1/2 & 1/2 \\
\hline
standard CH & 1/4 & 1/3 \\
\hline
\end{tabular}

\caption{Summary of coarsening exponents for different models and
base symmetries.} \label{tab_n}
\end{table}

{\it Discussion.}---We have found that the 2D coarsening exponent
for the standard GL and CH equations is the expected one
($n=\fra{1}{2}$ and $n=\fra{1}{3}$, respectively) for all symmetries
but the square one, which exhibits a slower coarsening . The
peculiar behaviour of square symmetry is at present not understood.
It must be noted, however, that if $\langle v^2\rangle\sim
q^{-\alpha}$ ($\alpha>1$, as occurs with the modified GL) then
$n=\fra{1}{2}$ for any symmetry. It seems thus that the peculiar
behavior of square symmetry is quite specific to the considered
equation, rather than general.
Nonetheless, a deeper understanding of this fact merits higher
attention in the future.

It is worthstressing once again that our approach allows to reduce
the study of dynamical properties (coarsening) to the behavior
of quantities ($D_{11},D_{12}$) depending on steady states
properties only.
Our analysis has been, for ease of presentation, exemplified on the
two classical equations, namely the GL and CH ones, but the method
can be applied to other equations.
In particular, our approach (as it is evident in Ref.~\cite{Politi_MisbahII})
does not require that the equation is derivable from a potential.

While for the GL and CH equations we could
extract analytically the coarsening exponent, it may prove necessary
that for  other equations there would be a need for numerical
solutions of the steady-states problem in order to evaluate the
diffusion constants. This is a quite simple task numerically, even
at higher dimensions (where full time-dependent studies are
difficult, or even unfeasible if the asymptotic regime is to be
ascertained).

The amplitude branch is  stable, in
contrast to the phase one when coarsening is to take place. There
are five different branches corresponding to different symmetries
(five Bravais lattices). The coarsening of the square branch is
slower (only for the classical GL and CH equations) than that of the
other symmetries. It is natural to expect the fastest growing
structure to prevail. Symmetries other than the square one offer,
for these classical equations,  a faster channel towards coarsening.

A particularly important question  concerns  relevance of slower coarsening peculiar to square symmetry. A possibility might be offered~\cite{Ciliberto} by the
Rayleigh-B\'enard convection, which may develop either rolls or
hexagons, with different local symmetries in different spatial
regions. These regions coarsen in time and the coarsening rate might
be different for domains of rolls and  hexagons.

We have investigated steady-state {\it periodic solutions}, while during coarsening,  even in 1D, no long range order is observed. The questions thus arises about importance  of periodic solutions.  Locally in space the amplitude is adiabatically slaved to the phase, so that (for a given wave vector) it reaches the quasi-steady solution.  If  steady-state solutions  were stable (no coarsening), then the system would generically choose the periodic solution (possibly with  defects), as is known for many pattern forming systems. Thus, having shown here that the periodic solutions are unstable with respect to phase fluctuations, we expect that the pattern should coarsen. One can not exclude, however, the existence of non-periodic (like disordered) stable solutions for nonlinear extended systems with an average wavelength which does not grow in time. We are not aware of any such  scenario, however.

Finally, it is noteworthy that the phase diffusion equations share
the same structures as those for the displacement field of
2D crystals~\cite{elasticity} (see also the Supplementary
Materials \cite{supp}).
Conservation law imposes
 $\partial_t \phi_i=-\partial _k J_{ik}$
 (this is the analogue of the dynamical equation in continuum media) where
 $J_{ik}$  is the
 phase current. $J_{ik}$ is proportional to the gradient
 of the phase, $\partial _l\phi _m$, and the proportionality coefficient is
 a rank four tensor, $J_{ik}=\Lambda_{kilm}\partial _l\phi _m$ (this is the analogue of Hooke's law).
  $\Lambda$ should be
 invariant under the symmetry group of the considered crystal. The
 phase equations has thus the same structure as the
 elasticity problem of crystals.

CM benefited from financial support from CNES.
PP acknowledges financial support from MIUR (PRIN 2007JHLPEZ).

\newpage

\widetext

\begin{center}
{\Large\bf Supplementary materials}
\vskip 0.5cm

{\large\bf Derivation of the phase equation from symmetry}
\end{center}
\vskip 1cm
Like in crystal elasticity, the phase equation has the form

$$\partial_t \psi_i=-\partial _k J_{ik}$$ where $J_{ik}$
is the phase
current which reads
$$J_{ik}=-\Lambda_{iklm}(\partial _l\psi _m+\partial _m\psi _l). $$
This is the analogue of Hooke's law. The minus sign in front $\Lambda$ expresses the fact that the current is opposite to the gradient.
Here we have written the current in a symmetrized form, since like
in elasticity, the deformation tensor (which is a measure of
distances in elasticity) is symmetric (the antisymmetric part corresponds to global rotation of the pattern, and is thus unimportant).

Thus the phase equation becomes
\begin{equation}
\partial_t \psi_i=-
  \Lambda_{iklm}\partial _{k} (\partial _l\psi _m+\partial _m\psi _l) \, ,
  \end{equation}
where $\Lambda$, which is the analogue of the  matrix defining the Lam\'e coefficients, should be
 invariant under the symmetry group of the considered crystal. Let us denote differentiation with respect to the variables $X_1$ and $X_2$ simply by $\partial _1$ and $\partial _2$, respectively.
 In a cubic crystal, as well as in a square lattice, only three
 components of $\Lambda$ are non zero (see Ref.[1]),
\begin{eqnarray}
&&\lambda_{1111}=\lambda_{2222}=\lambda_1\\
&&\lambda_{1122}=\lambda_{2211}=\lambda_2\\
&&\lambda_{1212}=\lambda_{2121}=\lambda_3 \, .
\end{eqnarray}
It follows that
\begin{equation}
\partial_t \psi_1=
  2\lambda_1 \partial _{11} \psi_1+ (2\lambda_2+\lambda_3) \partial
  _{12} \psi_2+\lambda_3 \partial
  _{22} \psi_{1} \, ,
  \end{equation}
  which has the form of our first equation (14). The same reasoning
  leads to the second equation (14). Regarding the first equation,
   term $\partial
  _{22} \psi_{2}$ is absent from symmetry, while in the second
  equation, $\partial
  _{11} \psi_{1}$ is absent.
\vskip 1cm

\noindent
[1]  L.D. Landau and E.M. Lifshitz, Theory of
Elasticity, Pergamon Press, Oxford (1987).


\begin{thebibliography}{99}

\bibitem{book_Ball}
P. Ball, {\it Branches, Shapes, Flow: Nature's Patterns: a Tapestry in Three Parts}
(Oxford University Press, 2009).

\bibitem{review_Bray}
A. J. Bray, Adv. Phys. {\bf 43}, 357 (1994).

\bibitem{coarsening_with_gravity}
M. Gratton and T. Witelski,
Phys. Rev. E 77, 016301 (2008).

\bibitem{Bray_Cugliandolo_PRL}
J. J. Arenzon, A. J. Bray, L. F. Cugliandolo, and A. Sicilia,
Phys. Rev. Lett. {\bf 98}, 145701 (2007).

\bibitem{Bray_Cugliandolo_PRE}
A. Sicilia, J. J. Arenzon, A. J. Bray, and L. F. Cugliandolo,
Phys. Rev. E {\bf 76}, 061116 (2007).

\bibitem{Lifshitz-Safran}
L. Gomez, E. Valles, and D. Vega,
Phys. Rev. Lett. {\bf 97}, 188302 (2006).

\bibitem{Watson}
S. J. Watson and S. A. Norris,
Phys. Rev. Lett. {\bf 96}, 176103 (2006).

\bibitem{Politi_MisbahI}
P. Politi and C. Misbah,
Phys. Rev. Lett. {\bf 92 }, 090601 (2004).

\bibitem{Kohn_Yan}
R. Kohn and X. Yan,
Comm. Pure Appl. Math. {\bf 56}, 1549 (2003).

\bibitem{driven_granular_media}
I. S. Aranson, B. Meerson, P. V. Sasorov, and V. M. Vinokur,
Phys. Rev. Lett. 88, 204301 (2002).

\bibitem{Bray_Rutenberg}
A. J. Bray and A. D. Rutenberg,
Phys. Rev. E {\bf 49}, R27 (1994).

\bibitem{Cross1995} M.C. Cross and D. Meiron, Phys. Rev. Lett. {\bf 75}, 2152 (1995).

\bibitem{Hohenberg_Halperin}
P. C. Hohenberg and B. I. Halperin,
Rev. Mod. Phys. {\bf 49}, 435 (1977).

\bibitem{Andreotti}
B. Andreotti et al., Nature {\bf 457}, 1120 (2009).

\bibitem{mounds}
F. Rabbering, H. Wormeester, F. Everts, and B. Poelsema,
Phys. Rev. B {\bf 79}, 075402 (2009).


\bibitem{book_Hoyle}
R. Hoyle, {\it Pattern formation: An introduction to methods}
(Cambridge University Press, 2006)

\bibitem{Politi_MisbahII}
P. Politi and C. Misbah, Phys. Rev. E {\bf 73}, 036133 (2006).

\bibitem{book_Nayfeh}
Ali H. Nayfeh, {\it Introduction to perturbation techniques}
(Wiley-Interscience, 1993).



\bibitem{Ciliberto}
Sergio Ciliberto, private communication.

\bibitem{elasticity}
A. E. H. Love, {\it A treatise on the mathematical theory of elasticity}
(Dover Publications, 1944).

\bibitem{supp} Supplementary materials.
\end{thebibliography}
\end{document}